# Record nighttime electric power generation at a density of 350 mW/m$^2$ via radiative cooling

Sid Assawaworrarit[1], Ming Zhou[1], Lingling Fan[1], and Shanhui Fan[1]

[1]*Department of Electrical Engineering, Ginzton Laboratory, Stanford University, Stanford, California, USA*

*\* To whom correspondence should be addressed. Email: shanhui@stanford.edu.*

## Abstract

The coldness of the universe is a thermodynamic resource that has largely remained untapped for renewable energy generation. Recently, a growing interest in this area has led to a number of studies with the aim to realize the potential of tapping this vast resource for energy generation. While the theoretical calculation based on thermodynamic principles places an upper limit in the power density at the level of 6000 mW/m$^2$, most experimental demonstrations so far result in much lower power density at the level of tens of mW/m$^2$. Here we demonstrate, through design optimization involving the tailoring of the thermal radiation spectrum, the minimization of parasitic heat leakage, and the maximum conversion of heat to electricity, an energy generation system harvesting electricity from the thermal radiation of the ambient heat to the cold universe that achieves a sustained power density of 350 mW/m$^2$. We further demonstrate a power density at the 1000 mW/m$^2$ level using an additional heat source or heat storage that provides access to heat at a temperature above ambient. Our work here shows that the coldness of the universe can be harvested to generate renewable energy at the power density level that approaches the established bound.

Amid the ongoing search for future sustainable source of energy, the cold universe emerges as an important thermodynamic resource that has been underutilized for renewable energy generation. The earth's atmosphere provides radiative access for an object on earth's surface to the cold universe due to its high transmission in the wavelength range between 8-13 um, coinciding with the peak of thermal radiation of objects at earth's temperature of ~300 K. A cold reservoir like the universe, with its temperature near absolute zero, thus serves an important role as a heat sink for thermal radiation from earth. While radiative cooling to the cold universe is a subject with a long history [1–5], the subject has received renewed interest in recent years [6–27] with its benefit of passive, electricity-free cooling amid growing demand in cooling needs. Recent years have also seen an increased interest in the use of the cold universe for energy generation on earth [28–33]. Due to its ability to generate power at night, such a technology is particularly attractive in off-grid electricity generation where other renewable sources like solar are absent at night and in the powering of remote sensors.

How much electricity can be generated on earth by radiative cooling from earth's surface at ~300 K to the universe? In a previous calculation [29], a thermal emitter with optimized emissivity spectrum connected to an ideal Carnot engine can produce ~6000 mW/m$^2$ of the generated power density – the power generated per unit area used to emit radiation – operating on earth's surface under a realistic atmosphere. Higher power density is possible thermodynamically but requires the use of non-reciprocal management of the thermal radiation which are more difficult [34]. Despite great promises, the generated power density demonstrated thus far remains low in the tens of milliwatts per square meter [28,29,32,33,31]. Improving the power density to reach 1000 mW/m$^2$ would make the technology more viable in meeting the energy demand of consumer applications. We recently demonstrated 100 mW/m$^2$ peak power density achieved by focusing on optimizing the sizing of components [33]. In this work, we optimize the design of the entire system in all aspects from the thermal radiation to thermal insulation and to electricity generation and experimentally demonstrate a new record performance.

We begin by introducing a simplified model of the heat transfer, previously used in our previous work [33], to calculate the power density that can be achieved with a particular design. (The full model is given in Methods.) This simplified model will elucidate the ways to realize a power density maximizing design.

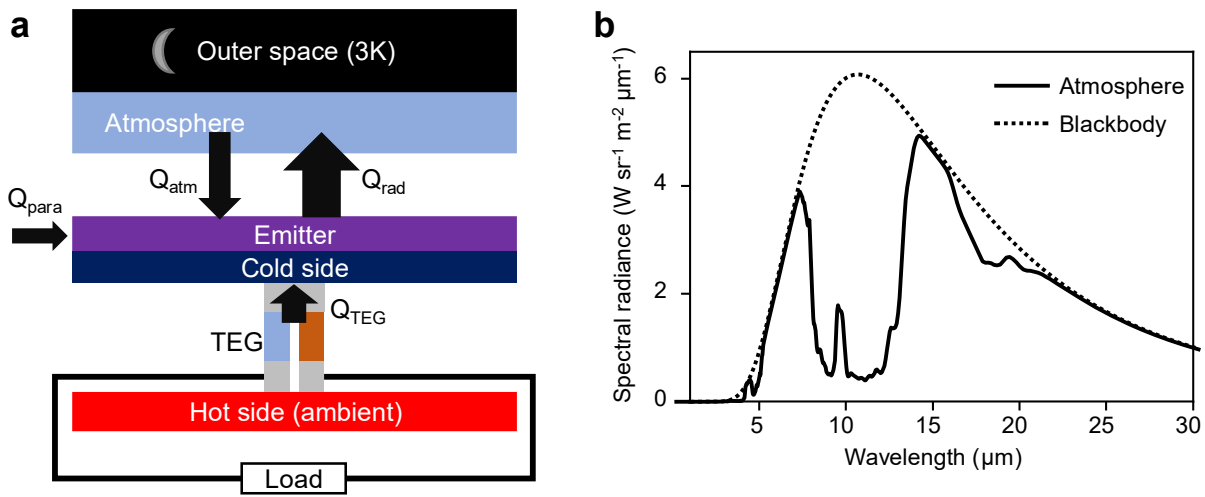

Figure 1: a, Schematic of a radiative cooling energy harvesting system. Solid black arrows indicate heat flows to and from the emitter. b, Spectral radiance of the downward radiation of the earth's atmosphere at normal incidence obtained from LOWTRAN mid-latitude winter model and that of a blackbody at 273 K for comparison.

Figure 1a shows the schematic of a radiative cooling energy harvesting system highlighting the relevant heat flows. A thermal emitter panel facing the night sky emits thermal radiation skyward ($Q_{rad}$) and simultaneously absorbs the downward radiation coming from the air mass that comprises the earth's atmosphere ($Q_{atm}$). Figure 1b shows an example of the spectral radiance of the atmosphere downward radiation using the LOWTRAN [35] mid-latitude winter model at normal incidence. The outline of the curve follows that of the blackbody spectrum at 273 K (dotted curve) except with the spectral content in the 8 – 13 um wavelength range substantially suppressed. This atmospheric transmission window plays an important role in radiative cooling. A thermal emitter designed to avoid absorbing the downward radiation from the atmosphere by, for instance, reflecting back the radiation achieves passive radiative cooling, that is, there is a net positive cooling radiative heat flow ($Q_{cool} = Q_{rad} - Q_{atm} > 0$) when the emitter is at the ambient temperature. This cooling heat flow increases with increasing emitter's temperature with the exact temperature dependence according to the law of thermal radiation (see Methods). Our simplified model treats this dependence on the emitter's temperature using a linear approximation around the ambient temperature, expressing the net cooling power as,

$$Q_{cool} \cong q_0 A + h_{rad} A (T_{emit} - T_{amb}).$$

where $A$ is the emitter's area; $T_{emit}$ and $T_{amb}$ are the temperatures of the emitter and the ambient, respectively; $q_0$ is the cooling power density when the emitter is at the ambient temperature; and $h_{rad}$ is the temperature coefficient of the cooling power density. The last two quantities, $q_0$ and $h_{rad}$, are functions of the emitter design.

This cooling is then used to run a generator: a thermoelectric generator (TEG) serving as a heat engine extracts useful work in the form of electrical energy from the heat flow from the ambient surroundings as a hot reservoir to the emitter as a cold reservoir. The TEG opens up a channel that draw heat from the ambient surroundings to the emitter in order to generate power. For a TEG, neglecting Joule heating, the heat flow can be expressed using a thermal resistance model as,

$$Q_{TEG} = \frac{T_{amb} - T_{emit}}{R_{TEG}} \quad (2)$$

where $R_{TEG}$ is the thermal resistance of the TEG. The maximum electric power generated by the TEG at load matching condition follows the quadratic dependence on the temperature difference across it, $p_{gen} = \alpha \Delta T^2$, where $\alpha$ is constant of a particular TEG, for small $\Delta T$. Finally, we lump together all other heat transfer channels between the emitter and the ambient into a parasitic term. This can come from convection or radiation to/from the back side. We write this as,

$$Q_{para} = h_{para} A (T_{amb} - T_{emit}). \quad (3)$$

Putting them all together, the power density can be calculated from the energy balance of the emitter: $Q_{cool} - Q_{TEG} - Q_{para} = 0$ to obtain,

$$w = \frac{p_{gen}}{A} = \frac{\alpha}{A} \left( \frac{q_0}{\left(h_{rad} + h_{para} + \frac{1}{AR_{TEG}}\right)} \right)^2. \quad (4)$$

From Eq. 4, it can be shown there is an area that maximizes the power density,

$$A^* = \frac{1}{R_{TEG}(h_{rad} + h_{para})}. \quad (5)$$

With this area, the power density is,

$$w(A = A^*) = \frac{\alpha_{TEG} R_{TEG}}{h_{rad} + h_{para}} \cdot \frac{q_0^2}{4}. \quad (6)$$

This expression (Eq. 6) reveals three components to the system-wide optimization. First, we look for the best performing TEG with the highest $\alpha_{TEG} R_{TEG}$ product. Second, we minimize parasitic heat transfer, $h_{para}$. Third, we optimize the design of the emitter which affects $q_0$ and $h_{rad}$ to maximize the power density.

Using a TEG with the highest possible $\alpha_{TEG} R_{TEG}$ product ensures that the TEG converts the heat flow through itself with the highest possible efficiency. These quantities can be obtained from manufacturers' datasheets. Alternatively, it can be shown that the product $\alpha_{TEG} R_{TEG}$ is proportional to the component $Z$ in $ZT$, the standard figure of merit for TEGs. Thus, the search for the best TEG is equivalent to finding one with the highest $ZT$ value in the expected operating temperature range of ~ 300K. We selected KELK KTGS066A00 TEG module with $\alpha_{TEG} R_{TEG} = 7 \times 10^{-5} \ K^{-1}$ obtained from the manufacturer's data sheet.

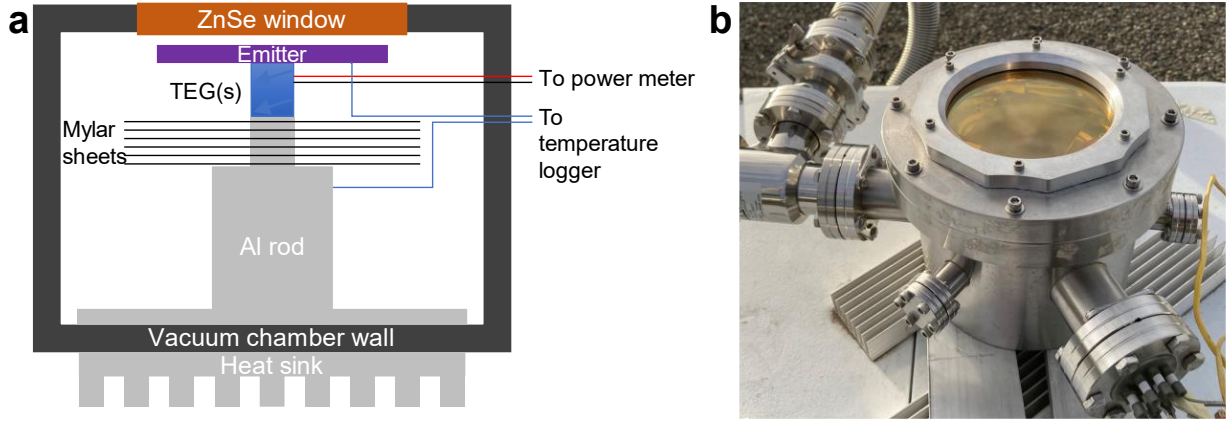

Figure 2: Design of our radiative cooling energy harvesting system. a, component diagram showing the arrangement inside the vacuum chamber of our setup. b, Photo of our setup.

Next, minimizing the parasitic heat transfer term, $h_{para}$, requires a high level of thermal insulation of the emitter and the cold side of the TEG. To approach the maximum power density requires $h_{para} \ll h_{rad}$ as evident in Eq. 6. Typical values for $h_{rad}$ ranges from 2-5 W/m²/K depending on the exact spectral response of the emitter. For an emitter fully exposed to air, open-air convection results in $h_{para}$ ~ 20 W/m²/K, significantly larger than $h_{rad}$. A foam insulation setup with transparent cover, as was used in many radiative cooling experiments [28,32,33], reduces $h_{para}$ down to ~ 5 W/m²/K, on the same order of magnitude as typical $h_{rad}$ values. To achieve significantly reduced $h_{para}$, we employ a vacuum setup. A previous experiment on radiative cooling to a very low temperature used a vacuum system and reported $h_{para}$ of ~ 0.1 - 0.3 W/m²/K [10]. This vacuum system uses a ZnSe top window with anti-reflection coating designed to maximize transmission at wavelengths centered at 10.6 μm. The window has over 80% transmission over the atmosphere transmission window of 8 – 13 μm, allowing the cooling radiation $Q_{rad}$ in this region to flow largely unimpeded. This vacuum setup still represents a factor of ~2 improvement in power density compared to a foam insulation one when both the reduced $h_{para}$ and the transmission of ZnSe are taken into account. We adapt this vacuum setup for energy harvesting. Figure 2 shows the component diagram and the photo of our setup.

Starting with a vacuum chamber with ZnSe window, we added an internal central aluminum rod with 1.5-in diameter to provide thermally conductive path from the TEG hot side to the bottom side of the vacuum chamber which functions as an ambient contact. An aluminum ring installed at the bottom locates the rod at the center of the vacuum chamber. To minimize parasitic heat transfer between the emitter and internal surfaces of the vacuum chamber, the top 1 cm of this rod is lathed down to 1 cm diameter, the diagonal dimension of the selected TEG, and a stack of 12 mylar discs spaced by nylon washers is installed on this top section. Mylar sheets are also plastered on all internal surfaces of the vacuum chamber to minimize thermal radiation that would otherwise heat up the emitter. Outside the vacuum chamber, we add large heat sinks underneath the vacuum chamber for ambient contact. Thermal paste is applied to all thermal conductive interfaces.

Finally, we optimize the design of thermal radiation of the emitter. For a thermal emitter to cool below the ambient in order to maximize the power density, the emitter should avoid absorbing the downward radiation from the atmosphere and, simultaneously, maximally emit its thermal radiation. For a reciprocal emitter, the Kirchoff's law of thermal radiation bounds the absorptivity to the emissivity, $\alpha(p, \lambda, \Omega) = \epsilon(p, \lambda, \Omega)$, for a given polarization ($p$), wavelength ($\lambda$), and angle ($\Omega$) as a result of the time reversal relationship between the absorption and emission processes [36]. The emitter's emissivity function $\epsilon(p, \lambda, \Omega)$ then modulates the $p, \lambda, \Omega$-channel for the radiative heat transfer between the emitter and the atmosphere. In our case of maximizing the power density, the optimum emissivity function is to have unity emissivity in a $p, \lambda, \Omega$-channel if doing so makes the outgoing radiation content in the channel larger than that absorbed from the atmosphere in the channel, and zero emissivity otherwise [29]. This generally translates to having unity emissivity in the atmosphere transmission window and zero outside.

To come up with a physically realizable emitter design with emissivity profile close to ideal, we base our emitter structure on a multilayer optical thin film design and use a memetic algorithm [37] for optimization. The emissivity/absorptivity of a multilayer structure can be calculated from the complex refractive indices of the constituent materials using the transfer matrix method [38]. The optimization algorithm begins by creating multilayer structures with the constituent layer materials randomly chosen from a materials library with randomized layer thicknesses. Then the algorithm uses a combination of evolutionary algorithm and local optimization to fine-tune the material and thickness for each layer in the multilayer structure that best achieve a given objective function. Such an approach has previously been shown to yield an emitter design with near-ideal emissivity profile for their given objective function, at least when evaluated using computation based on the tabulated refractive index library [29,37]. However, since the optical properties of fabricated thin films can depend on the precise film deposition techniques and conditions as reported by others [39] and experienced in our early trials, we narrow our search space to those materials whose optical properties we can reliably fabricate in our facility: Si, SiO, $SiO_2$, and $SiN_x$ for dielectric films and Al for metallic film. These candidate materials should be sufficient to build an emitter that achieves high power density: Si film is transparent in the thermal wavelengths; SiO, $SiO_2$, and $SiN_x$ films have absorption peaks in the atmosphere transmission window; and Al film has high reflection.

We ran the optimization using the power density (Eq. 6) as the objective function for up to 10 maximum number of layers. We selected the solution with three layers: 600 nm $SiN_x$ on 500 nm Si on an Al reflector layer. Higher numbers of layers bring diminishing return in terms of power density at the expense of much longer fabrication time. We fabricated the films on top of a silicon wafer using sputtering deposition in the Stanford Nanofabrication Facility (Methods). Figure 3a shows the photo of the fabricated sample and its emissivity spectra measured using a spectrometer (Nicolet 6700 with Harrick's Seagull variable angle accessory). The spectra show high emissivity in the atmosphere transmission window as desired. The calculated maximum power density is 330 mW/m$^2$.

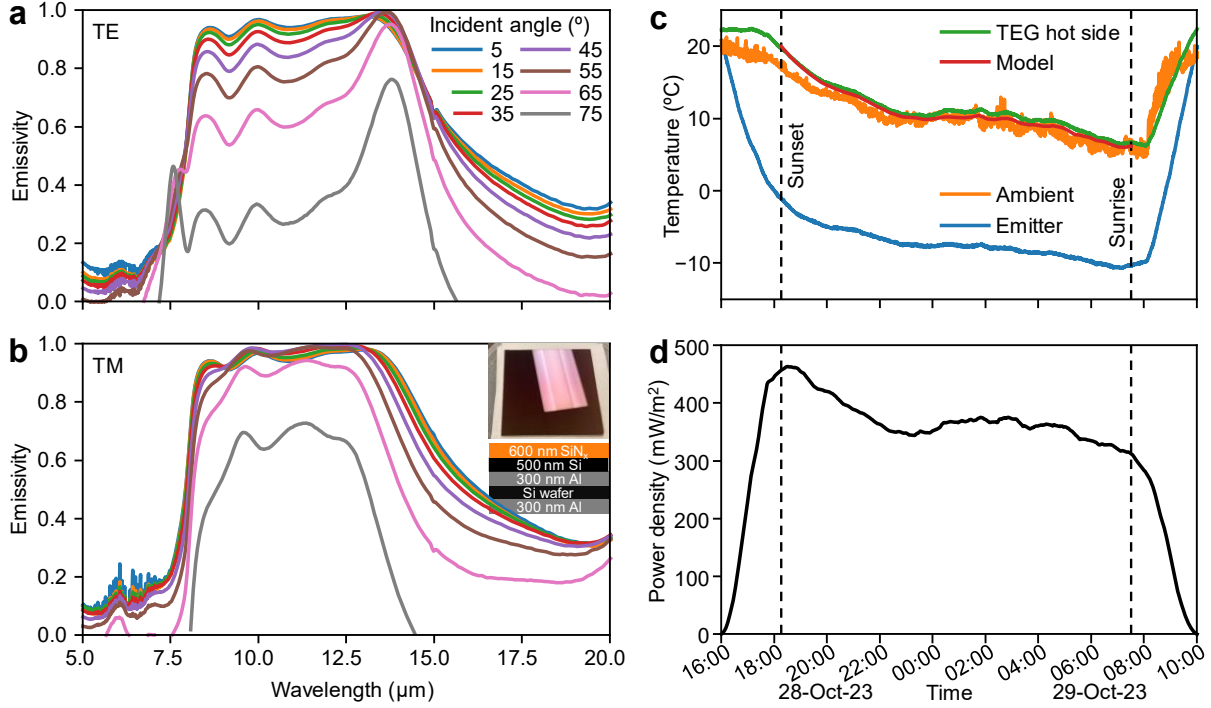

Figure 3: Results. a,b Emissivity spectra for TE (a) and TM (b) polarization of our fabricated emitter. Inset in b: photo and layer structure of our fabricated emitter. c,d Measured power density (c) and temperature (d) achieved using our setup.

To demonstrate this power density experimentally, we install the fabricated thermal emitter in the prepared vacuum chamber setup shown in Fig. 2. The area that maximizes the power density (Eq. 5) for the chosen TEG is 153 cm$^2$. Due to the small size of the vacuum chamber, we stacked up five identical TEG modules vertically and connect their outputs electrically in series. Stacking $n$ identical TEG modules vertically creates a combined TEG module with $R_{TEG} \to nR_{TEG}$ and $\alpha \to \alpha/n$ [33]. Stacking five TEG modules thus allow us to demonstrate the same power density with a fifth of the original optimal emitter area that fits inside our vacuum chamber. We cut the fabricated emitter to 5.6 cm x 5.6 cm and installed it above the TEG stack. The central rod was designed to position the emitter just underneath the top ZnSe window with around 4 mm gap between them. Figure 2b shows the finished experimental setup which we located on the roof top of Stanford's Electrical Engineering building. We pumped down the vacuum chamber to ~$10^{-5}$ Tor using a vacuum pump (Pfeiffer HiCube 80). We used a data logger (Omega RDXL6SD) and type-K thermocouples to monitor the temperatures of the emitter, the central rod inside the vacuum chamber, and the ambient. We used a source measurement unit (Keithley 2635B) running a maximum power point tracking program to measure the generated power from the series-connected TEGs. We also monitored the ambient temperature, humidity, and wind conditions using a weather station (Acurite Iris 5-in-1 weather sensor). Supplementary Figure 1 shows the full experimental setup.

We observed the power generation and other measurements from late October 2023 to mid-January 2024. On nights with clear sky, we regularly measured power generation density at 300 mW/m$^2$ and above, and cooling of the emitter's temperature at 14-18 ºC below ambient. Figure 3b shows power generation density and temperature measurements on the night of October 29, 2023. (Supplementary Fig. 2 shows results from other days.) Here the power density peaks at over 450 mW/m$^2$ around sunset before falling down to settle at around 350 mW/m2 level. We attribute this pre-sunset peak to the effect of thermal storage in the vacuum chamber that receives heat from sunlight during daytime and retains its heat after sunset. Hence, the temperature of the vacuum chamber, which connects to the hot side of the TEG, is higher that of the ambient surroundings during daytime and, as the night cools down, remains elevated for a while after sunset. Also, the ambient temperature continually declines into the night, accentuating the difference. Figure 3c shows the TEG hot side temperature, measured at the internal central rod close to its interface with the TEG, is higher than the ambient temperature by around 3 ºC at sunset and takes approximately two hours to converge to the ambient temperature. A thermal storage model can explain this heat holding effect of the vacuum chamber. Before sunset, sunlight heats up the vacuum chamber to an above-ambient temperature. After sunset, with no solar heat input, the temperature of the vacuum chamber relaxes to that of the ambient with a relaxation time constant, $\tau_{VC}$. With $\tau_{VC} = 40$ minutes (red curve in Fig. 3c), this thermal storage model agrees well with the measured temperature. This $\tau_{VC}$ is in line with our expectation given the mass of the vacuum chamber and the thermal design of the ambient contact. This thermal storage effect is known inflate the power generation density value to be higher than it would be if the hot side of the TEG were actually at ambient temperature, potentially leading to overly optimistic power density number being quoted [40]. The recommended steps of using sunshade and large heat sinks help reduce but do not entirely eliminate the thermal storage effect in our case due to the large thermal mass of the vacuum chamber. Accordingly, in this work, we take the number 350 mW/m$^2$, the level that the experimental setup achieved and consistently sustained throughout the night when the temperature of the vacuum chamber closely follows the ambient level, to be the true power density achieved from energy harvesting using the ambient surroundings as a heat source for thermal radiation to the cold universe.

While such an emphasis for measuring power density when the hot side of the TEG is at ambient temperature is done in the interest of fair comparison, in practice, an additional thermal source or thermal storage effect of terrestrial objects such as buildings, soils, and bodies of water can be used to boost the power density in similar energy harvesting setups [41,42]. Here, we mimic the effect of thermal storage by adding a heat pad to the base of the vacuum chamber. We connect a proportional-integral-derivative (PID) controller (Omega CN7523) to the heat pad allowing us to heat the vacuum chamber to a set temperature to simulate the use of a hotter-than-ambient temperature from a heat storage with large thermal mass. Figures 4a and 4b show the temperatures and measured power density when the controller is set to 18 °C. With access to this above-ambient heat source, the power density reaches 700 mW/m$^2$, a substantial gain compared to using the ambient heat alone. Figure 4c shows the calculated power density as a function of the ambient and heat source temperatures. Given access to a heat storage with temperature around 20 ºC above ambient, our calculation shows that power density at a level of 1000 mW/m$^2$ can be achieved.

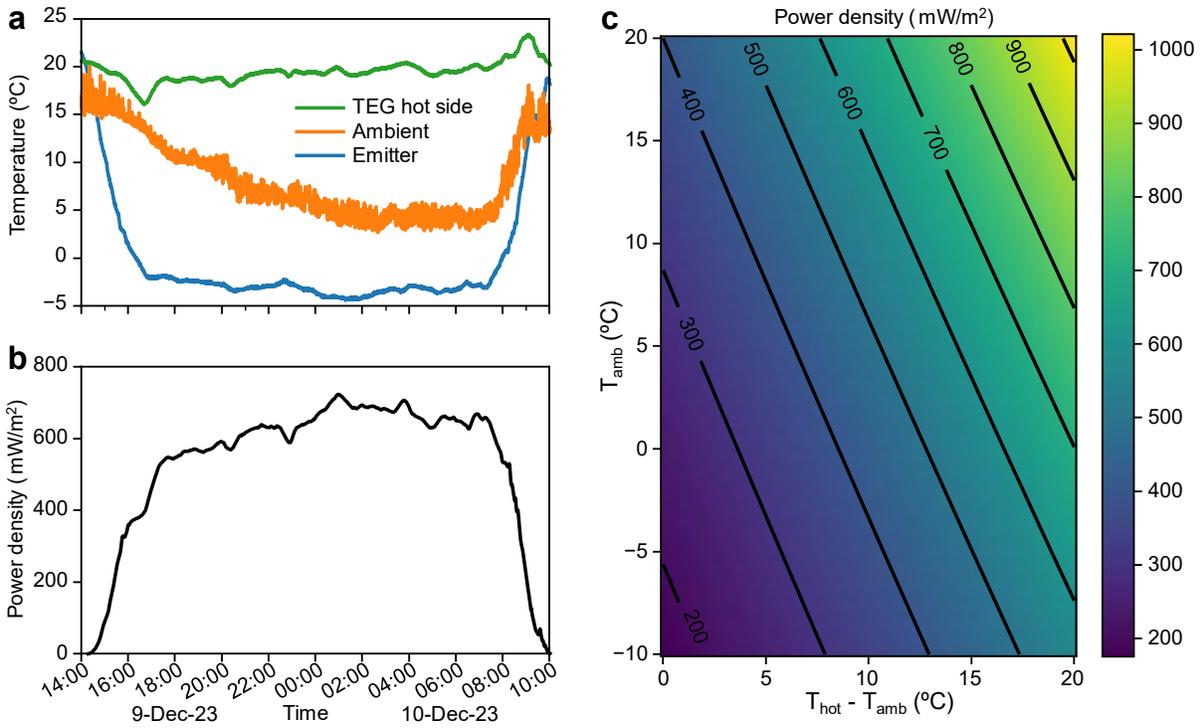

Figure 4: Using thermal storage to increase the power density. a,b Measured temperature (a) and power density (b) when the hot side of the TEG is maintained at around 18 ºC (b). c, Calculated power density as a function of ambient temperature and the temperature differential of the TEG hot side above ambient.

In summary, the coldness of the universe is the thermodynamic resource that can be used to generate energy on earth. Here we use the thermal radiation that flows out to the universe to cool down an emitter to below ambient temperature and use a thermoelectric generator to harvest energy from the temperature difference between the emitter and the ambient. We consider the system model to optimize all its components for maximum power density: the thermoelectric generator and sizing; the parasitic loss; and the emitter's spectral response. We experimentally demonstrate 350 mW/m$^2$ power density when harvesting from the ambient heat alone. We further demonstrate 1000 mW/m$^2$-level power density using a heat storage effect. Our work here shows that with careful design, the power density of energy harvesting using the coldness of the universe can approach the established theoretical limit.

This work is supported by the Stanford Strategic Energy Alliance, and by the U. S. Department of Energy (Grant No. DE-FG02-07ER46426). Part of this work was performed at the Stanford Nanofabrication Facility (SNF), which is supported by the National Science Foundation under award ECCS-2026822.

Methods

Full thermal model: The thermal radiation model follows the treatment in [10] which considers the transmission of thermal radiation through the ZnSe window with the following expressions for the radiative cooling heat flow terms:

$$Q_{rad}(T_{emit}) = A\int d\Omega \cos\theta \int d\lambda I_{bb}(T_{emit},\lambda)\epsilon_{emit}(\lambda,\Omega)\frac{t_w(\lambda) + \alpha_w(\lambda)}{1 + r_w(\lambda)[\epsilon_{emit}(\Omega) - 1]},$$

where $\theta(\Omega)$ is the (solid) angle a photon leaves the emitter; $I_{bb}(T,\lambda) = \frac{2hc^2}{\lambda^5\left[\exp\left(\frac{hc}{\lambda k_B T}\right)-1\right]}$ is the spectral intensity of a blackbody at temperature $T$; $\lambda$ is the wavelength, $\epsilon_{emit}$ is the emissivity of the emitter; $t_w, \alpha_w, r_w$ are the transmission, absorption, and reflection, respectively, of the ZnSe window; and $T_{emit}$ is the emitter temperature.

$$Q_{atm}(T_{amb}) = A\int d\Omega \cos\theta \int d\lambda I_{bb}(T_{amb},\lambda)\epsilon_{emit}(\lambda,\Omega)\frac{t_w(\lambda)\epsilon_{atm}(\lambda) + \alpha_w(\lambda)}{1 + r_w(\lambda)[\epsilon_{emit}(\Omega) - 1]}$$

where $T_{amb}$ is the ambient temperature and $\epsilon_{atm}$ is the emissivity of the atmosphere. Here we model the atmosphere's downward radiation as a blackbody at $T_{amb}$ modulated by its emissivity $\epsilon_{atm}$ obtained from the Lowtran mid-latitude winter model.

In the case of energy harvesting from ambient temperature only, we solve the energy balance $Q_{cool} - Q_{TEG} - Q_{para} = 0$ where $Q_{cool} = Q_{rad} - Q_{atm}$ for the $T_{emit}$ and then use the temperature reduction to calculate the power density. In the case of additional thermal source or thermal storage, we use $Q_{TEG} = \frac{T_{hot} - T_{emit}}{R_{TEG}}$ in place of Eq. 2 where $T_{hot}$ is the temperature of the thermal source or storage.

Emitter fabrication:

We used the Stanford Nanofabrication Facility (SNF) for our emitter fabrication. Starting with a 4-in silicon substrate, we deposit 300 nm Al, 500 nm Si, and 600 nm SiNx sequentially on top of the substrate. We also deposit 300 nm Al on the back side as an added measure to reflect radiation coming from below. We used Lesker's PVD 200 sputtering physical vapor deposition machine and used deposition targets obtained from Lesker. For SiN$_x$ film, we determined the optical constants of our developed films for the purpose of our optimization by fabricating additional samples of SiNx films on Al film [43].

# Record nighttime electric power generation at a density of 350 mW/m$^2$ via radiative cooling: Supplementary Information

Sid Assawaworrarit[1], Ming Zhou[1], Lingling Fan[1], and Shanhui Fan[1]

[1]*Department of Electrical Engineering, Ginzton Laboratory, Stanford University, Stanford, California, USA*

*\* To whom correspondence should be addressed. Email: shanhui@stanford.edu.*

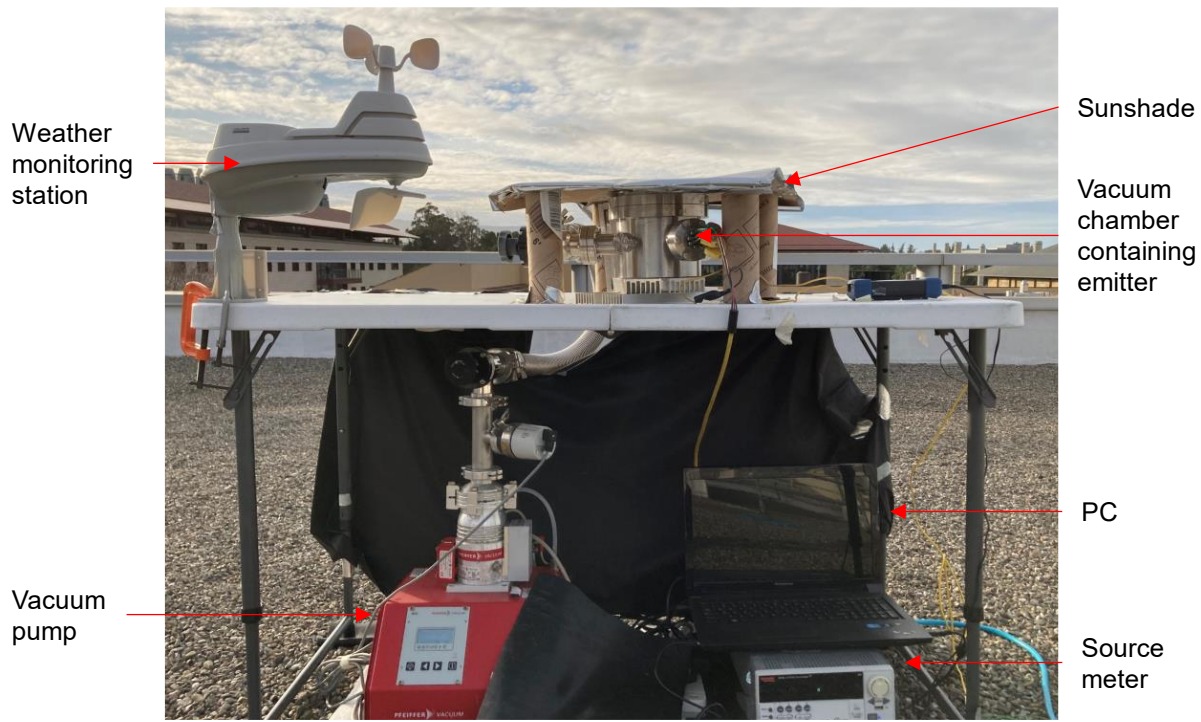

Supplementary Figure 1: Experimental setup.

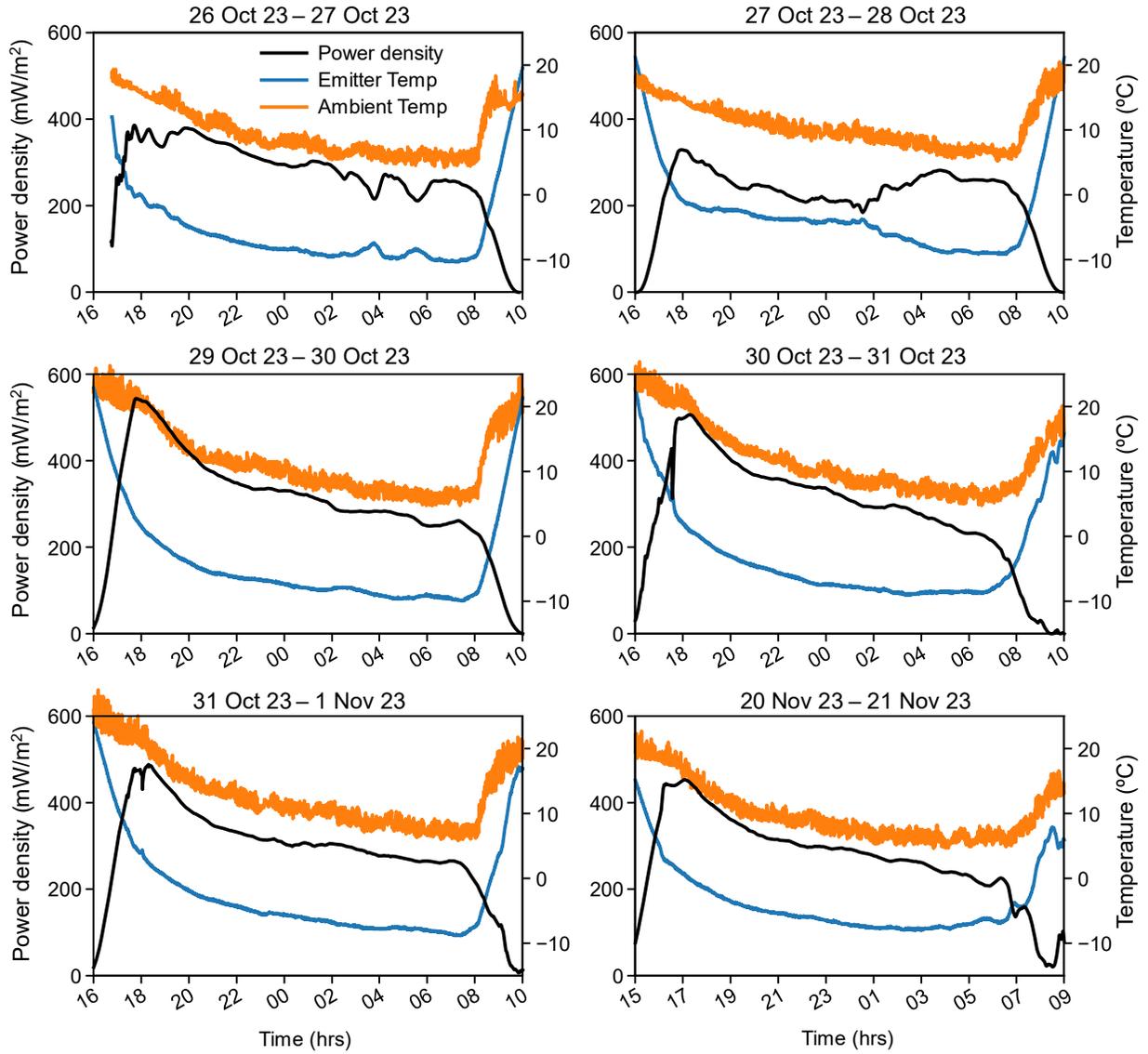

Supplementary Figure 2: Additional experimental results from other observation nights. Note: local daylight saving time ends on Nov 5, 2023.